\documentclass[aps,pre,amsmath,10pt,twocolumn,superscriptaddress]{revtex4-1}
\usepackage{graphicx}
\usepackage{color}
\usepackage{bm}
\usepackage{braket}

\begin{document}
\title{Machine Learning Potential Repository}
\author{Atsuto \surname{Seko}}
\email{seko@cms.mtl.kyoto-u.ac.jp}
\affiliation{Department of Materials Science and Engineering, Kyoto University, Kyoto 606-8501, Japan}
\date{\today}

\begin{abstract}
This paper introduces a machine learning potential repository that includes Pareto optimal machine learning potentials. 
It also shows the systematic development of accurate and fast machine learning potentials for a wide range of elemental systems.
As a result, many Pareto optimal machine learning potentials are available in the repository from a website \cite{MachineLearningPotentialRepository}.
Therefore, the repository will help many scientists to perform accurate and fast atomistic simulations.
\end{abstract}

\maketitle

\section{Introduction}

Machine learning potential (MLP) has been increasingly required to perform crystal structure optimizations and large-scale atomistic simulations more accurately than with conventional interatomic potentials.
Therefore, many recent studies have proposed a number of procedures to develop MLPs and have shown their applications
\cite{
Lorenz2004210,
behler2007generalized,
bartok2010gaussian,
behler2011atom,
han2017deep,
258c531ae5de4f5699e2eec2de51c84f,
PhysRevB.96.014112,
PhysRevB.90.104108,
PhysRevX.8.041048,
PhysRevLett.114.096405,
PhysRevB.95.214302,
PhysRevB.90.024101,
PhysRevB.92.054113,
PhysRevMaterials.1.063801,
Thompson2015316,
wood2018extending,
PhysRevMaterials.1.043603,
doi-10.1137-15M1054183,
PhysRevLett.120.156001,
podryabinkin2018accelerating,
GUBAEV2019148,
doi:10.1063/1.5126336}.
Simultaneously, MLPs themselves are necessary for their users to perform accurate atomistic simulations. 
Therefore, the development and distribution of MLPs for a wide range of systems should be useful, similarly to the conventional interatomic potentials distributed in several repositories \cite{interatomicPotentialRepository,KIMproject}.

This study demonstrates an MLP repository available from a website \cite{MachineLearningPotentialRepository}. 
The MLP repository includes Pareto optimal MLPs with different trade-offs between accuracy and computational efficiency because they are conflicting properties  and there is no single optimal MLP \cite{PhysRevB.99.214108,hernandez2019fast,doi:10.1021/acs.jpca.9b08723}.
This study develops the repository by performing systematic density functional theory (DFT) calculations for approximately 460,000 structures and by combining them with existing DFT datasets in the literature \cite{doi:10.1063/1.5027283,PhysRevB.99.214108}. 
Polynomial-based potential energy models \cite{doi:10.1063/1.5027283,PhysRevB.99.214108} and their revisions are then systematically applied to the construction of MLPs for a wide range of elemental systems.
Although the present version of the repository does not contain MLPs for multicomponent systems, the repository will be gradually updated. 
Moreover, a user package that combines MLPs in the repository with atomistic simulations using the \textsc{lammps} code \cite{lammps} is also available on a website \cite{LammpsMLIPpackage}.


\section{Potential energy models}
This section shows structural features and potential energy models used for developing MLPs in the repository.
Given cutoff radius $r_c$ from atom $i$ in a structure, the short-range part of the total energy for the structure may be decomposed as
\begin{equation}
E = \sum_i E^{(i)},
\end{equation}
where $E^{(i)}$ denotes the contribution of atom $i$ or the atomic energy.
The atomic energy is then given by a function of invariants for the O(3) group \cite{bartok2013representing,PhysRevB.99.214108} as
\begin{equation}
E^{(i)} = F \left( d_1^{(i)}, d_2^{(i)}, \cdots \right),
\end{equation}
where $d_n^{(i)}$ denotes an invariant derived from order parameters representing the neighboring atomic density of atom $i$.
In the context of MLPs, invariants $\{d_n^{(i)}\}$ can be called ``structural features''.
Also, a number of functions are useful as function $F$ to represent the relationship between the invariants and the atomic energy, such as
artificial neural network models
\cite{
Lorenz2004210,
behler2007generalized,
behler2011atom,
han2017deep,
258c531ae5de4f5699e2eec2de51c84f,
PhysRevB.96.014112}, 
Gaussian process models
\cite{
bartok2010gaussian,
PhysRevB.90.104108,
PhysRevX.8.041048,
PhysRevLett.114.096405,
PhysRevB.95.214302},
and linear models
\cite{
PhysRevB.90.024101,
PhysRevB.92.054113,
PhysRevMaterials.1.063801,
Thompson2015316,
wood2018extending,
PhysRevMaterials.1.043603,
doi-10.1137-15M1054183}.
In the repository, linear models are explained as function $F$, which are shown in Sec. \ref{Sec-Models}.

\subsection{Structural features}
A systematic procedure to derive a set of structural features that can control the accuracy and computational efficiency of MLPs (e.g., \cite{bartok2013representing,PhysRevB.99.214108}) plays an essential role in automatically generating fast and accurate MLPs. 
Therefore, the repository employs systematic sets of structural features derived from order parameters representing the neighboring atomic density in terms of a basis set. 
They are classified into a set of structural features derived only from radial functions and a set of structural features derived from radial and spherical harmonic functions.

A pairwise structural feature is expressed as
\begin{equation}
d_{n0}^{(i)} = \sum_{j \in {\rm neighbor}} f_n(r_{ij}),
\end{equation}
where $r_{ij}$ denotes the distance between atoms $i$ and $j$.
The repository adopts a finite basis set of Gaussian-type radial functions given by
\begin{equation}
f_{n}(r)=\exp\left[-\beta_n(r-r_n)^{2}\right] f_c(r),
\end{equation}
where $\beta_n$ and $r_n$ denote parameters.
Cutoff function $f_c$ ensures smooth decay of the radial function, and the repository employs a cosine-based cutoff function expressed as 
\begin{eqnarray}
f_c(r) = \left\{
\begin{aligned}
& \frac{1}{2} \left[ \cos \left( \pi \frac{r}{r_c} \right) + 1\right] & (r \le r_c)\\
& 0 & (r > r_c)
\end{aligned}
\right ..
\end{eqnarray}

Another structural feature is a linearly independent polynomial invariant of the O(3) group, which is generated from order parameters representing the neighboring atomic density in terms of spherical harmonics.
A $p$th-order polynomial invariant for a given radial number $n$ and a given set of angular numbers $\set{l_1,l_2,\cdots,l_p}$ is defined by a linear combination of products of $p$ order parameters, expressed as
\begin{widetext}
\begin{equation}
\label{Eqn-invariant-form}
d_{nl_1l_2\cdots l_p, (\sigma)}^{(i)} = \sum_{\set{m_1,m_2,\cdots, m_p}} C^{l_1l_2\cdots l_p, (\sigma)}_{m_1m_2\cdots m_p} a_{nl_1m_1}^{(i)} a_{nl_2m_2}^{(i)} \cdots a_{nl_pm_p}^{(i)},
\end{equation}
\end{widetext}
where $a_{nlm}^{(i)}$ denotes the order parameter of component $nlm$ representing the neighboring atomic density of atom $i$.
The order parameters for atom $i$ in a given structure are approximately calculated from its neighboring atomic density regardless of the orthonormality of the radial functions \cite{PhysRevB.99.214108} as
\begin{equation}
a_{nlm}^{(i)} = \sum_{j \in \rm {neighbor}} f_n(r_{ij}) Y_{lm}^* (\theta_{ij}, \phi_{ij}),
\label{EquationOrderParameters}
\end{equation}
where $(r_{ij}, \theta_{ij}, \phi_{ij})$ denotes the spherical coordinates of neighboring atom $j$ centered at the position of atom $i$.
A coefficient set $\set{C^{l_1l_2\cdots l_p, (\sigma)}_{m_1m_2\cdots m_p}}$ is determined by using a group-theoretical projection operator method \cite{el-batanouny_wooten_2008}, ensuring that the linear combination of Eqn. (\ref{Eqn-invariant-form}) is invariant for arbitrary rotation \cite{PhysRevB.99.214108}.
In terms of fourth- and higher-order polynomial invariants, there exist multiple invariants that are linearly independent for most of the set $\set{l_1,l_2,\cdots,l_p}$. 
Therefore, they are distinguished by index $\sigma$ if necessary.

\subsection{Energy models with respect to structural features}
\label{Sec-Models}
The repository uses polynomial functions as function $F$ representing the relationship between the atomic energy and a given set of structural features, $D = \set{d_1,d_2,\cdots}$.
The polynomial functions with regression coefficients $\set{w}$ are given as follows.
\begin{eqnarray}
F_1 \left(D\right) &=& \sum_i w_i d_i \nonumber \\
F_{2,\rm pow} \left(D\right) &=& \sum_{i} w_{ii} d_i d_i \nonumber \\
F_2 \left(D\right) &=& \sum_{\{i,j\}} w_{ij} d_i d_j \\
F_{3,\rm pow} \left(D\right) &=& \sum_{i} w_{iii} d_i d_i d_i \nonumber \\
F_3 \left(D\right) &=& \sum_{\{i,j,k\}} w_{ijk} d_i d_j d_k \nonumber \\
& \vdots & \nonumber
\end{eqnarray}

A potential energy model is identified with a combination of the polynomial functions and structural features.
The repository introduces the following six potential energy models.
When a set of pairwise structural features is described as $D_{\rm pair}^{(i)} = \set{d_{n0}^{(i)}}$, the first model (\texttt{model type = 1, feature type = pair}) is composed of powers of the pairwise structural features as
\begin{equation}
E^{(i)} = F_1\left(D_{\rm pair}^{(i)} \right)
+ F_{2,\rm pow} \left(D_{\rm pair}^{(i)} \right)
+ F_{3,\rm pow} \left(D_{\rm pair}^{(i)} \right)
+ \cdots,
\end{equation}
which is measured from the energy of the isolated state of atom $i$.
This model was proposed in Refs. \onlinecite{PhysRevB.90.024101} and \onlinecite{PhysRevB.92.054113}.
The second model (\texttt{model type = 2, feature type = pair}) is a polynomial of the pairwise structural features with their cross terms, expressed as
\begin{equation}
E^{(i)} = F_1\left(D_{\rm pair}^{(i)} \right)
+ F_{2} \left(D_{\rm pair}^{(i)} \right)
+ F_{3} \left(D_{\rm pair}^{(i)} \right)
+ \cdots.
\end{equation}
This model can be regarded as a natural extension of embedded atom method (EAM) potentials as demonstrated in Ref. \onlinecite{PhysRevMaterials.1.063801}.

The other four models are derived from the polynomial invariants of Eqn. (\ref{Eqn-invariant-form}).
When a set of the polynomial invariants is expressed by the union of sets of $p$th-order polynomial invariants as
\begin{equation}
D^{(i)} = D_{\rm pair}^{(i)} \cup D_2^{(i)}
\cup D_3^{(i)} \cup D_4^{(i)} \cup \cdots,
\end{equation}
where 
\begin{eqnarray}
D_2^{(i)} &=& \set{d_{nll}^{(i)}} \nonumber \\
D_3^{(i)} &=& \set{d_{nl_1l_2l_3}^{(i)}} \\ 
D_4^{(i)} &=& \set{d_{nl_1l_2l_3l_4,(\sigma)}^{(i)}}, \nonumber
\end{eqnarray}
the third model (\texttt{model type = 1, feature type = invariants}) is expressed as
\begin{equation}
E^{(i)} = F_1 \left(D^{(i)} \right)
+ F_{2,\rm pow} \left(D^{(i)} \right)
+ F_{3,\rm pow} \left(D^{(i)} \right)
+ \cdots.
\label{Eqn-linear-polynomial}
\end{equation}
This model consists of the powers of the polynomial invariants. 
A linear polynomial form of the polynomial invariants, $E^{(i)} = F_1 \left(D^{(i)} \right)$, which was proposed in Ref. \onlinecite{PhysRevB.99.214108}, is included in the third model.
Note that a linear polynomial model with up to third-order invariants, expressed by
\begin{equation}
E^{(i)} = F_1 \left( D_{\rm pair}^{(i)} \cup D_2^{(i)} \cup D_3^{(i)} \right),
\end{equation}
is regarded as a spectral neighbor analysis potential (SNAP) \cite{Thompson2015316}.

The fourth model (\texttt{model type = 2, feature type = invariants}) is given by a polynomial of the polynomial invariants as
\begin{equation}
E^{(i)} = F_1 \left(D^{(i)} \right)
+ F_{2} \left(D^{(i)} \right)
+ F_{3} \left(D^{(i)} \right)
+ \cdots .
\end{equation}
A quadratic polynomial model of the polynomial invariants up to the third order, expressed as
\begin{equation}
E^{(i)} = F_1 \left( D_{\rm pair}^{(i)} \cup D_2^{(i)} \cup D_3^{(i)} \right) +  F_2 \left( D_{\rm pair}^{(i)} \cup D_2^{(i)} \cup D_3^{(i)} \right),
\end{equation}
is regarded as a quadratic SNAP \cite{doi:10.1063/1.5017641}.

The fifth model (\texttt{model type = 3, feature type = invariants}) is the sum of a linear polynomial form of the polynomial invariants and a polynomial of pairwise structural features, described as
\begin{equation}
E^{(i)} = f_1 \left( D^{(i)} \right) 
+ f_2 \left( D_{\rm pair}^{(i)} \right)
+ f_3 \left( D_{\rm pair}^{(i)} \right) 
+ \cdots .
\end{equation}
The sixth model (\texttt{model type = 4, feature type = invariants}) is the sum of a linear polynomial form of the polynomial invariants and a polynomial of pairwise structural features and second-order polynomial invariants.
This is written as
\begin{equation}
E^{(i)} = f_1 \left( D^{(i)} \right) 
+ f_2 \left( D_{\rm pair}^{(i)} \cup D_2^{(i)} \right) 
+ \cdots .
\end{equation}

\section{Datasets}
Training and test datasets are generated from prototype structures, i.e., structure generators.
The repository uses two sets of structure generators for elemental systems.
One is composed of face-centered cubic (fcc), body-centered cubic (bcc), hexagonal close-packed (hcp), simple cubic (sc), $\omega$, and $\beta$-tin structures, which was employed in Ref. \onlinecite{doi:10.1063/1.5027283}. 
Hereafter, structures generated from the structure generator set are denoted by ``dataset 1''.
The other is composed of prototype structures reported in the Inorganic Crystal Structure Database (ICSD) \cite{bergerhoff1987crystal}, which aims to generate a wide variety of structures.
For elemental systems, only prototype structures composed of single elements with zero oxidation state are chosen from the ICSD.
The total number of the structure generators is 86. 
A list of structure generators can be found in the Appendix of Ref. \onlinecite{PhysRevB.99.214108}.
Hereafter, structures generated from the second set are denoted by ``dataset 2''.

Given a structure generator, the atomic positions and lattice constants of the structure generator are fully optimized by DFT calculation to obtain its equilibrium structure.
Then, a new structure is constructed by random lattice expansion, random lattice distortion, and random atomic displacements into a supercell of the structure generator.
For a given parameter $\varepsilon$ controlling the degree of lattice expansion, lattice distortion, and atomic displacements, the lattice vectors of the new structure $\bm{A'}$ and the fractional coordinates of an atom in the new structure $\bm{f'}$ are given as
\begin{eqnarray}
\bm{A'} &=& \bm{A} + \varepsilon\bm{R} \\
\bm{f'} &=& \bm{f} + \varepsilon \bm{A'}^{-1} \bm{\eta},
\end{eqnarray}
where the $(3\times3)$ matrix $\bm{R}$ and the three-dimensional vector $\bm{\eta}$ are composed of uniform random numbers ranging from $-1$ to 1.
Matrix $\bm{A}$ and vector $\bm{f}$ represent the lattice vectors of the original supercell and the fractional coordinates of the atom in the original supercell, respectively.

For each elemental system, datasets 1 and 2 are composed of 3,000 and 10,000 structures, respectively, in addition to the equilibrium structures of the structure generators.
Dataset 1 was developed in Ref. \onlinecite{doi:10.1063/1.5027283}, whereas dataset 2, except for the case of elemental aluminum, is developed in this study.
Each of the datasets is then randomly divided into a training dataset and a test dataset.
In the repository, datasets 1 and 2 are available for 31 and 47 elements, respectively.
This means that the repository contains MLPs developed from a total of 567,228 DFT calculations.

DFT calculations were performed using the plane-wave-basis projector augmented wave method \cite{PAW1} within the Perdew--Burke--Ernzerhof exchange-correlation functional \cite{GGA:PBE96} as implemented in the \textsc{vasp} code \cite{VASP1,VASP2,PAW2}.
The cutoff energy was set to 300 eV.
The total energies converged to less than 10$^{-3}$ meV/supercell.
The atomic positions and lattice constants of the structure generators were optimized until the residual forces were less than 10$^{-2}$ eV/\AA.

\section{Model coefficient estimation}

Coefficients of a potential energy model are estimated from all the total energies, forces, and stress tensors included in a training dataset.
Given a potential energy model, therefore, the predictor matrix and observation vector are simply written in a submatrix form as
\begin{equation}
\bm{X} =
\begin{bmatrix}
\bm{X}_{\rm energy} \\
\bm{X}_{\rm force} \\
\bm{X}_{\rm stress} \\
\end{bmatrix}
,\qquad \bm{y} =
\begin{bmatrix}
\bm{y}_{\rm energy} \\
\bm{y}_{\rm force} \\
\bm{y}_{\rm stress} \\
\end{bmatrix}.
\end{equation}
The predictor matrix $\bm{X}$ is divided into three submatrices, $\bm{X}_{\rm energy}$, $\bm{X}_{\rm force}$, and $\bm{X}_{\rm stress}$, which contain structural features and their polynomial contributions to the total energies, the forces acting on atoms, and the stress tensors of structures in the training dataset, respectively.
The observation vector $\bm{y}$ also has three components, $\bm{y}_{\rm energy}$, $\bm{y}_{\rm force}$, and $\bm{y}_{\rm stress}$, which contain the total energy, the forces acting on atoms, and the stress tensors of structures in the training dataset, respectively, obtained from DFT calculations.
Using the predictor matrix and the observation vector, coefficients of a potential energy model are estimated by linear ridge regression.

In the case of dataset 2 for elemental aluminum, the training data has 9,086, 1,314,879, and 54,516 entries for the energy, the force, and the stress tensor, respectively.
Therefore, the predictor matrix $\bm{X}$ has a size of $(1,378,481, n_{\rm coeff})$, where $n_{\rm coeff}$ denotes the number of coefficients of the potential energy model and ranges from $10$ to $32,850$ in the potential energy models of the repository.

\section{Pareto optimality}
The accuracy and computational efficiency of the present MLP strongly depend on the given input parameters.
They are 
(1) the cutoff radius, 
(2) the type of structural features,
(3) the type of potential energy model, 
(4) the number of radial functions, 
(5) the polynomial order in the potential energy model, 
and
(6) the truncation of the polynomial invariants, i.e, the maximum angular numbers of spherical harmonics, $\{l_{\rm max}^{(2)}, l_{\rm max}^{(3)}, \cdots, l_{\rm max}^{(p_{\rm max})}\}$ and the polynomial order of invariants, $p_{\rm max}$.
Therefore, a systematic grid search is performed for each system to find their optimal values. 
The input parameters used for developing MLPs can be found in the repository.

However, it is difficult to find the optimal set of parameters because the accuracy and computational efficiency of an MLP are conflicting properties whose trade-off should be optimized,  as pointed out in Ref. \onlinecite{PhysRevB.99.214108}.
In this multiobjective optimization problem involving several conflicting objectives, there is no single optimal solution but a set of alternatives with different trade-offs between the accuracy and the computational efficiency.
In such a case, Pareto optimal points can be optimal solutions with different trade-offs \cite{branke2008multiobjective}.
Therefore, the repository contains all Pareto optimal MLPs for each system and each dataset.

\section{MLPs in repository}

\begin{figure}[tbp]
\includegraphics[clip,width=\linewidth]{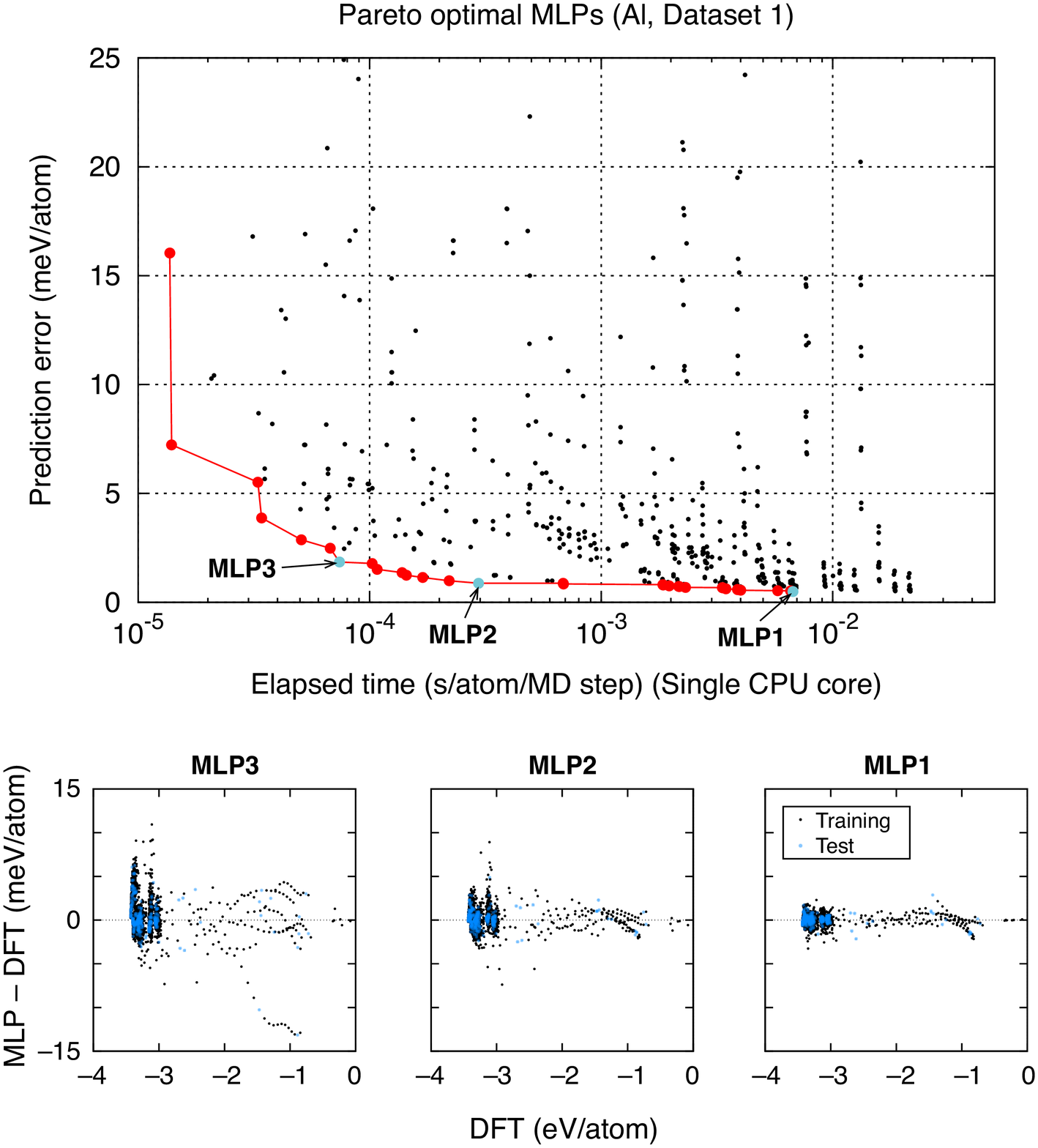}
\caption{
Distribution of MLPs in a grid search to find optimal parameters controlling the accuracy and the computational efficiency of the MLP for elemental Al. 
The elapsed time for a single point calculation is estimated using a single core of Intel\textregistered\ Xeon\textregistered\ E5-2695 v4 (2.10 GHz).
The red closed circles show the Pareto optimal points of the distribution obtained using a non-dominated sorting algorithm.
The cyan closed circles indicate the MLP with the lowest prediction error and two Pareto optimal MLPs with higher computational cost performance.
The distribution of the prediction errors for dataset 1 is also shown.
}
\label{mlip-db-2020:pareto-Al}
\end{figure}

\begin{table}[tbp]
\begin{ruledtabular}
\caption{
Model parameters of MLP1, MLP2, and MLP3 for elemental Al.
}
\label{mlip-db-2020:parameter-Al}
\begin{tabular}{lccc}
& MLP1 & MLP2 & MLP3 \\
\hline
Number of coefficients & 2410 & 1770 & 815 \\
Feature type & Invariants & Pair & Pair \\
Cutoff radius & 12.0 & 12.0 & 8.0 \\
Number of radial functions & 20 & 20 & 15\\
Model type & 3 & 2 & 2 \\
Polynomial order (function $F$) & 3 & 3 & 3 \\
Polynomial order (invariants) & 4 & $-$ & $-$ \\
$\set{l_{\rm max}^{(2)}, l_{\rm max}^{(3)},\cdots}$ & [4,4,2] & $-$ & $-$ \\
\end{tabular}
\end{ruledtabular}
\end{table}

\begin{figure*}[tbp]
\includegraphics[clip,width=0.85\linewidth]{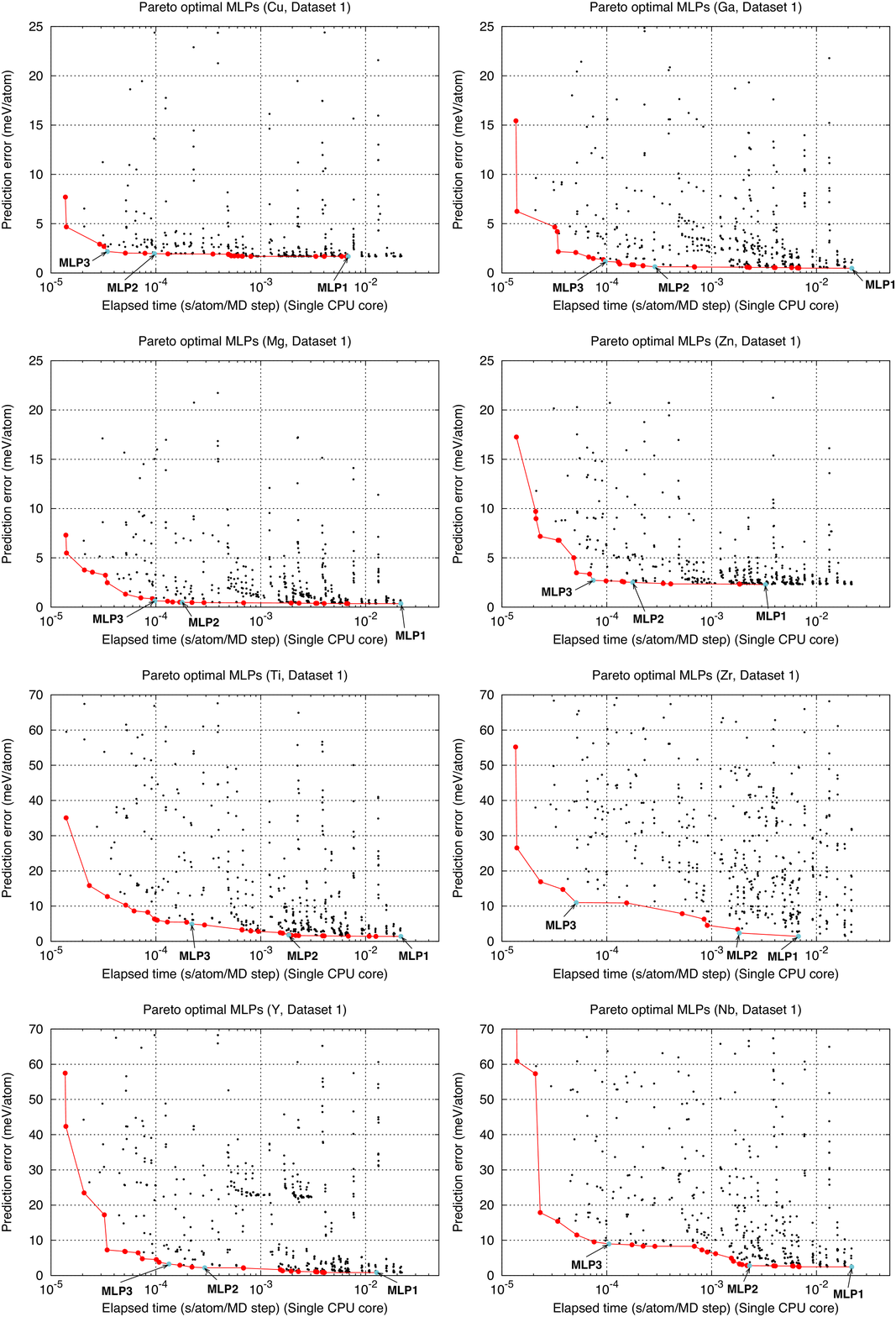}
\caption{
Distribution of MLPs in a grid search for elemental Cu, Ga, Mg, Zn, Ti, Zr, Y, and Nb.
The closed red circles show the Pareto optimal points of the distribution.
}
\label{mlip-db-2020:pareto-element}
\end{figure*}

Figure \ref{mlip-db-2020:pareto-Al} shows the prediction error and the computational efficiency of the Pareto optimal MLPs developed from dataset 1 for elemental Al.
Figure \ref{mlip-db-2020:pareto-element} also shows the Pareto optimal MLPs for elemental Cu, Ga, Mg, Zn, Ti, Zr, Y, and Nb.
The prediction error is estimated using the root mean square (RMS) error of the energy for the test dataset.
The computational efficiency is estimated using the elapsed time to compute the energy, the forces and the stress tensors of a structure with 284 atoms.
In Figs. \ref{mlip-db-2020:pareto-Al} and \ref{mlip-db-2020:pareto-element}, the elapsed time is normalized by the number of atoms because it is proportional to the number of atoms as shown later.
The behavior of the relationship between the prediction error and the computational efficiency for the other systems can be found in the repository.

Users of the repository can choose an appropriate MLP from the Pareto optimal ones according to their targets and purposes.
The MLP with the lowest prediction error is denoted by ``MLP1'', whereas two Pareto optimal MLPs showing higher computational cost performance are denoted by ``MLP2'' and ``MLP3''.
As can be seen in Figs. \ref{mlip-db-2020:pareto-Al} and \ref{mlip-db-2020:pareto-element}, MLP2 and MLP3 exhibit high computational efficiency without significantly increasing the prediction error.
This study introduces simple scores to evaluate the computational cost performance from the elapsed time $t$ with the unit of ms/atom/step and the prediction error $\Delta E$ with the unit of meV/atom.
MLP2 and MLP3 with higher computational cost performance minimize $t + \Delta E$ and $10t + \Delta E$, respectively. 

Figure \ref{mlip-db-2020:pareto-Al} shows the distribution of the prediction errors for structures in dataset 1.
Table \ref{mlip-db-2020:parameter-Al} also lists the values of the model parameters of MLP1, MLP2, and MLP3.
This information for the other Pareto optimal MLPs and the other systems can be found in the repository.

\begin{table*}[tbp]
\begin{ruledtabular}
\caption{
Prediction error and computational efficiency of MLPs constructed from dataset 1 for 31 elemental systems.
The normalized elapsed time for a single point calculation, the RMS error for the energy, and the RMS error for the force are denoted by $t$ (s/atom/step), $\Delta E$ (meV/atom), and $\Delta f$ (eV/\AA), respectively. 
MLP1 shows the lowest prediction error of $\Delta E$.
MLP2 and MLP3 show the lowest values of $t + \Delta E$ and $10t + \Delta E$, respectively. 
}
\label{mlip-db-2020:error1}
\begin{tabular}{c|ccc|ccc|ccc}
& \multicolumn{3}{c}{MLP1} & \multicolumn{3}{c}{MLP2} & \multicolumn{3}{c}{MLP3} \\
& $t$ & $\Delta E$ & $\Delta f$ & $t$ & $\Delta E$ & $\Delta f$ & $t$ & $\Delta E$ & $\Delta f$ \\
\hline
Ag & 10.71 & 1.9 & 0.004 & 0.07 & 2.0 & 0.008 & 0.03 & 2.2 & 0.011 \\
Al & 6.74 & 0.5 & 0.006 & 0.30 & 0.9 & 0.014 & 0.07 & 1.8 & 0.016 \\
Au & 21.65 & 0.5 & 0.006 & 0.66 & 0.7 & 0.012 & 0.05 & 1.8 & 0.027 \\
Ba & 5.79 & 1.0 & 0.005 & 0.30 & 1.2 & 0.011 & 0.10 & 1.8 & 0.013 \\
Be & 12.55 & 1.1 & 0.019 & 1.54 & 2.0 & 0.026 & 0.13 & 5.5 & 0.043 \\
Ca & 15.83 & 1.0 & 0.004 & 0.07 & 1.1 & 0.011 & 0.07 & 1.1 & 0.011 \\
Cd & 2.02 & 4.6 & 0.016 & 0.13 & 5.0 & 0.011 & 0.05 & 5.3 & 0.018 \\
Cr & 12.64 & 2.8 & 0.061 & 2.31 & 3.6 & 0.070 & 0.80 & 5.5 & 0.082 \\
Cs & 6.60 & 0.5 & 0.001 & 0.16 & 0.5 & 0.002 & 0.12 & 0.6 & 0.001 \\
Cu & 6.82 & 1.7 & 0.004 & 0.10 & 2.0 & 0.011 & 0.03 & 2.2 & 0.013 \\
Ga & 21.73 & 0.5 & 0.006 & 0.29 & 0.6 & 0.014 & 0.10 & 1.2 & 0.015 \\
Hf & 21.72 & 0.9 & 0.039 & 1.85 & 1.4 & 0.051 & 0.18 & 4.3 & 0.103 \\
Hg & 4.01 & 0.8 & 0.004 & 0.23 & 1.0 & 0.008 & 0.07 & 1.2 & 0.010 \\
In & 21.67 & 0.5 & 0.005 & 0.22 & 0.7 & 0.014 & 0.07 & 1.2 & 0.014 \\
K & 18.53 & 0.4 & 0.000 & 0.09 & 0.5 & 0.001 & 0.09 & 0.5 & 0.001 \\
Li & 6.89 & 0.1 & 0.001 & 0.13 & 0.2 & 0.003 & 0.03 & 0.7 & 0.005 \\
Mg & 21.71 & 0.4 & 0.002 & 0.18 & 0.5 & 0.006 & 0.10 & 0.6 & 0.007 \\
Mo & 21.69 & 2.4 & 0.065 & 2.16 & 3.3 & 0.078 & 0.15 & 9.3 & 0.138 \\
Na & 21.68 & 0.2 & 0.001 & 0.10 & 0.2 & 0.001 & 0.05 & 0.4 & 0.002 \\
Nb & 21.65 & 2.4 & 0.048 & 2.18 & 2.8 & 0.058 & 0.10 & 9.0 & 0.127 \\
Rb & 12.48 & 0.5 & 0.001 & 0.12 & 0.5 & 0.001 & 0.09 & 0.7 & 0.001 \\
Sc & 21.71 & 0.7 & 0.017 & 0.22 & 2.6 & 0.048 & 0.18 & 3.0 & 0.049 \\
Sr & 21.53 & 0.5 & 0.003 & 0.22 & 0.7 & 0.008 & 0.13 & 0.8 & 0.009 \\
Ta & 21.70 & 1.6 & 0.056 & 0.91 & 3.3 & 0.071 & 0.22 & 8.5 & 0.126 \\
Ti & 21.67 & 1.4 & 0.035 & 1.85 & 1.8 & 0.047 & 0.13 & 5.5 & 0.100 \\
Tl & 21.66 & 0.5 & 0.005 & 0.29 & 0.8 & 0.012 & 0.10 & 1.6 & 0.014 \\
V & 6.95 & 2.2 & 0.048 & 1.09 & 3.1 & 0.058 & 0.07 & 8.5 & 0.129 \\
W & 21.69 & 2.9 & 0.080 & 2.32 & 3.9 & 0.092 & 0.14 & 12.0 & 0.177 \\
Y & 12.77 & 0.8 & 0.016 & 0.29 & 2.2 & 0.040 & 0.13 & 3.3 & 0.044 \\
Zn & 3.27 & 2.3 & 0.008 & 0.18 & 2.5 & 0.011 & 0.07 & 2.7 & 0.014 \\
Zr & 6.71 & 1.4 & 0.044 & 1.84 & 2.4 & 0.055 & 0.05 & 11.0 & 0.116 \\
\end{tabular}
\end{ruledtabular}
\end{table*}

\begin{table*}[tbp]
\begin{ruledtabular}
\caption{
Prediction error and computational efficiency of MLPs constructed from dataset 2 for 47 elemental systems.
}
\label{mlip-db-2020:error2}
\begin{tabular}{c|ccc|ccc|ccc}
& \multicolumn{3}{c}{MLP1} & \multicolumn{3}{c}{MLP2} & \multicolumn{3}{c}{MLP3} \\
& $t$ & $\Delta E$ & $\Delta f$ & $t$ & $\Delta E$ & $\Delta f$ & $t$ & $\Delta E$ & $\Delta f$ \\
\hline
Ag & 18.51 & 1.1 & 0.019 & 0.76 & 1.3 & 0.033 & 0.28 & 2.5 & 0.035 \\
Al & 28.69 & 1.8 & 0.033 & 1.11 & 3.0 & 0.040 & 0.28 & 7.3 & 0.061 \\
As & 23.39 & 5.1 & 0.125 & 1.98 & 8.5 & 0.144 & 0.58 & 13.5 & 0.178 \\
Au & 23.49 & 3.1 & 0.028 & 0.75 & 3.9 & 0.035 & 0.28 & 7.0 & 0.056 \\
Ba & 36.80 & 0.7 & 0.013 & 1.44 & 1.9 & 0.021 & 0.13 & 4.6 & 0.034 \\
Be & 8.69 & 3.8 & 0.078 & 1.10 & 5.7 & 0.088 & 0.28 & 11.8 & 0.132 \\
Bi & 23.45 & 2.8 & 0.130 & 2.06 & 4.6 & 0.121 & 0.62 & 9.4 & 0.166 \\
Ca & 23.66 & 0.4 & 0.006 & 0.93 & 1.1 & 0.013 & 0.17 & 3.5 & 0.030 \\
Cd & 23.41 & 0.7 & 0.011 & 0.75 & 1.8 & 0.018 & 0.28 & 3.2 & 0.026 \\
Cr & 13.95 & 6.7 & 0.221 & 2.59 & 8.3 & 0.226 & 0.35 & 18.2 & 0.324 \\
Cs & 37.17 & 0.4 & 0.001 & 0.22 & 0.6 & 0.002 & 0.09 & 0.9 & 0.003 \\
Cu & 23.20 & 8.2 & 0.022 & 0.28 & 8.8 & 0.034 & 0.10 & 9.5 & 0.051 \\
Ga & 23.38 & 1.1 & 0.028 & 1.12 & 1.8 & 0.039 & 0.27 & 4.5 & 0.044 \\
Ge & 23.45 & 2.7 & 0.058 & 1.11 & 5.0 & 0.067 & 0.58 & 7.6 & 0.076 \\
Hf & 13.65 & 4.2 & 0.121 & 2.51 & 6.0 & 0.137 & 0.75 & 8.6 & 0.148 \\
Hg & 23.61 & 3.6 & 0.014 & 0.75 & 4.7 & 0.020 & 0.13 & 6.5 & 0.038 \\
In & 23.59 & 0.7 & 0.016 & 0.93 & 1.2 & 0.021 & 0.28 & 2.8 & 0.028 \\
Ir & 22.64 & 9.0 & 0.251 & 3.19 & 10.8 & 0.260 & 0.75 & 16.8 & 0.295 \\
K & 7.10 & 0.1 & 0.001 & 0.22 & 0.4 & 0.002 & 0.08 & 0.7 & 0.003 \\
La & 37.02 & 2.5 & 0.057 & 1.96 & 3.8 & 0.069 & 0.69 & 6.4 & 0.079 \\
Li & 23.59 & 0.2 & 0.004 & 0.75 & 0.9 & 0.010 & 0.10 & 1.7 & 0.019 \\
Mg & 23.62 & 0.3 & 0.006 & 0.75 & 0.8 & 0.009 & 0.14 & 2.9 & 0.029 \\
Mo & 18.21 & 7.3 & 0.211 & 3.48 & 8.6 & 0.226 & 0.75 & 15.6 & 0.266 \\
Na & 3.20 & 1.5 & 0.196 & 0.71 & 2.0 & 0.197 & 0.17 & 2.7 & 0.226 \\
Nb & 36.63 & 6.5 & 0.182 & 2.55 & 7.6 & 0.183 & 0.75 & 11.8 & 0.212 \\
Os & 14.15 & 10.2 & 0.304 & 3.21 & 11.4 & 0.300 & 0.75 & 18.8 & 0.348 \\
P & 23.76 & 7.3 & 0.176 & 1.98 & 9.8 & 0.186 & 1.11 & 11.8 & 0.192 \\
Pb & 23.98 & 1.2 & 0.028 & 0.94 & 2.4 & 0.037 & 0.17 & 5.0 & 0.057 \\
Pd & 14.51 & 2.6 & 0.073 & 0.99 & 4.0 & 0.080 & 0.28 & 7.6 & 0.096 \\
Pt & 23.19 & 5.3 & 0.137 & 1.11 & 7.3 & 0.148 & 0.72 & 8.2 & 0.156 \\
Rb & 36.57 & 0.0 & 0.000 & 0.13 & 0.5 & 0.002 & 0.07 & 0.9 & 0.004 \\
Re & 37.20 & 9.8 & 0.274 & 1.98 & 13.5 & 0.291 & 0.71 & 18.4 & 0.320 \\
Rh & 13.78 & 6.4 & 0.186 & 1.98 & 8.5 & 0.192 & 0.71 & 12.6 & 0.217 \\
Ru & 23.43 & 8.5 & 0.234 & 3.19 & 9.9 & 0.237 & 0.75 & 16.4 & 0.279 \\
Sb & 23.50 & 3.4 & 0.120 & 2.00 & 6.0 & 0.411 & 0.75 & 8.7 & 0.124 \\
Sc & 23.92 & 3.0 & 0.211 & 1.98 & 4.0 & 0.234 & 0.75 & 5.9 & 0.135 \\
Si & 23.58 & 4.1 & 0.077 & 1.11 & 7.2 & 0.088 & 0.75 & 8.9 & 0.095 \\
Sn & 23.45 & 1.7 & 0.036 & 1.12 & 3.5 & 0.049 & 0.58 & 5.5 & 0.061 \\
Sr & 11.80 & 0.7 & 0.007 & 0.76 & 1.6 & 0.014 & 0.18 & 3.2 & 0.022 \\
Ta & 22.94 & 6.5 & 0.190 & 3.17 & 7.7 & 0.195 & 0.75 & 12.3 & 0.221 \\
Ti & 13.20 & 4.4 & 0.143 & 1.98 & 6.4 & 0.146 & 0.69 & 9.2 & 0.163 \\
Tl & 24.14 & 0.8 & 0.015 & 0.72 & 2.2 & 0.023 & 0.15 & 5.0 & 0.060 \\
V & 14.23 & 6.4 & 0.188 & 2.54 & 8.4 & 0.196 & 0.71 & 12.3 & 0.228 \\
W & 22.71 & 8.3 & 0.247 & 3.17 & 9.8 & 0.254 & 0.99 & 14.7 & 0.286 \\
Y & 36.55 & 2.6 & 0.050 & 1.98 & 3.9 & 0.062 & 0.71 & 6.7 & 0.070 \\
Zn & 23.62 & 1.1 & 0.017 & 0.99 & 1.9 & 0.024 & 0.27 & 4.6 & 0.038 \\
Zr & 14.57 & 5.9 & 0.130 & 0.82 & 9.0 & 0.139 & 0.75 & 9.1 & 0.140 \\
\end{tabular}
\end{ruledtabular}
\end{table*}

Tables \ref{mlip-db-2020:error1} and \ref{mlip-db-2020:error2} list the prediction error and the computational efficiency of MLPs for each elemental system obtained from datasets 1 and 2, respectively.
MLP2 and MLP3 exhibit high computational efficiency while avoiding a significant increase of the prediction error.
Therefore, MLP2 and MLP3 can be regarded as better potentials than MLP1 for most practical purposes.

\begin{figure}[tbp]
\includegraphics[clip,width=0.9\linewidth]{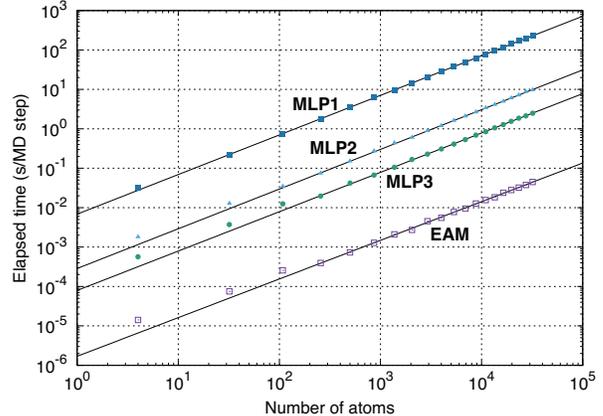}
\caption{
Dependence of the computational time required for a single point calculation on the number of atoms.
The elapsed time is measured using a single core of Intel\textregistered\ Xeon\textregistered\ E5-2695 v4 (2.10 GHz).
}
\label{mlip-db-2020:natoms-time}
\end{figure}

Figure \ref{mlip-db-2020:natoms-time} shows the elapsed times of single point calculations for structures with up to 32,000 atoms using the EAM potential \cite{PhysRevB.35.7423}, MLP1, MLP2, and MLP3 for elemental Al.
Structures were made by the expansion of the fcc conventional unit cell with a lattice constant of 4 \AA.
As can be seen in Fig. \ref{mlip-db-2020:natoms-time}, linear scaling with respect to the number of atoms is achieved in all the MLPs.
Although the performance for only three MLPs is shown here, the other MLPs also exhibit linear scaling with respect to the number of atoms.
Therefore, the computational time required for a calculation of $n_{\rm step}$ steps for a structure with $n_{\rm atom}$ atoms can be estimated as $ t \times n_{\rm atom} \times n_{\rm step}$, where $t$ is the elapsed time per atom for a single point calculation listed in the repository.

\section{Conclusion}
An MLP repository developed by a systematic application of the procedure to obtain Pareto optimal MLPs has been demonstrated in this paper.
In particular, MLPs with high computational cost performance, showing high computational efficiency without increasing the prediction error, are useful for most practical purposes.
Currently, many Pareto optimal MLPs are available in the repository from the website, and the number of MLP entries in the repository is continuously increasing.
Therefore,  the repository should be useful in performing accurate and fast atomistic simulations.

\begin{acknowledgments}
This work was supported by a Grant-in-Aid for Scientific Research (B) (Grant Number 19H02419) and a Grant-in-Aid for Scientific Research on Innovative Areas (Grant Number 19H05787) from the Japan Society for the Promotion of Science (JSPS).
\end{acknowledgments}

\bibliography{mlip-db}

\end{document}